\documentclass[prc,showpacs,showkeys,superscriptaddress,nofootinbib,twocolumn,floatfix]{revtex4-1}
\usepackage{amsfonts}
\usepackage{float}
\usepackage{placeins}
\usepackage{graphicx}
\usepackage[hidelinks]{hyperref}
\usepackage{color,amsmath,amssymb,bm}
\usepackage{tensor}

\usepackage[normalem]{ulem}  

\renewcommand{\sout}{\bgroup \color{red} \ULdepth=-.5ex \ULset}

\def\blfootnote{\xdef\@thefnmark{}\@footnotetext}

\newcommand{\beq}{\begin{equation}}
\newcommand{\eeq}{\end{equation}}
\newcommand{\bea}{\begin{eqnarray}}
\newcommand{\eea}{\end{eqnarray}}

\begin{document}


\title{Evolution of $\Lambda$ polarization in the hadronic phase of heavy-ion collisions}

\author{Yifeng Sun}
\affiliation{Department of Physics and Astronomy, University of Catania, Via S. Sofia 64, 1-95125 Catania, Italy}
\affiliation{Laboratori Nazionali del Sud, INFN-LNS, Via S. Sofia 62, I-95123 Catania, Italy}

\author{Zhen Zhang}
\email{zhangzh275@mail.sysu.edu.cn}
\affiliation{Sino-French Institute of Nuclear Engineering and Technology, Sun Yat-sen University, Zhuhai 519082, China}

\author{Che Ming Ko}
\affiliation{Cyclotron Institute and Department of Physics and Astronomy, Texas A$\&$M University, College Station, Texas 77843-3366, USA}


\author{Wenbin Zhao}
\affiliation{Department of Physics and Astronomy, Wayne State University, Detroit, Michigan 48201, USA}

\date{\today}

\begin{abstract}
Using the AMPT + MUSIC+UrQMD hybrid model, we study the global and local spin polarizations of $\Lambda$ hyperons as functions of the freeze-out temperature of the spin degree of freedom in the hadronic phase of Au+Au collisions at $\sqrt{s_{NN}}=19.6$ GeV. Including contributions from both the thermal vorticity and thermal shear of the hadronic matter, we find that, with the spin freeze-out temperature dropping from the hadronization temperature of 160 MeV to 110 MeV at the kinetic freeze-out, both the global and local spin polarizations of $\Lambda$ hyperons due to the thermal vorticity decrease by a factor of two, while those due to the thermal shear decrease quickly and become negligibly small at 140 MeV. Our results suggest the importance of understanding the dynamical evolution  of the spin degree of freedom in the hadronic stage in relativistic heavy-ion collisions.
\end{abstract}

\maketitle

\section{Introduction}

In non-central heavy ion collisions at relativistic energies, a large amount of orbital angular momentum from the two colliding nuclei are transferred to the produced quark-gluon plasma (QGP), creating thus the most vortical fluid of the order of  $10^{21}-10^{22}$ s$^{-1}$ in known physical systems~\cite{STAR:2017ckg,PhysRevC.93.064907,PhysRevC.94.044910,PhysRevC.87.034906,PhysRevC.90.021904,PhysRevC.88.061901,PhysRevC.92.014906}.  Due to their spin-orbit interactions, quarks in the QGP and $\Lambda$ hyperons formed after the hadronization become  polarized along the direction of the total orbital angular momentum~\cite{PhysRevLett.96.039901,PhysRevC.77.044902}, as observed in experiments by the STAR Collaboration ~\cite{STAR:2017ckg,PhysRevLett.126.162301}.   Because of the non-uniformity of the vorticity field, local structures in the $\Lambda$ spin polarization have also been found both in theoretical studies~\cite{PhysRevLett.117.192301,PhysRevLett.120.012302,PhysRevC.98.024905} and in experimental measurements~\cite{PhysRevLett.123.132301}. Although various theoretical models, such as those based on the hydrodynamic approach~\cite{PhysRevC.88.034905,Karpenko2017,PhysRevC.94.054907,PhysRevC.95.031901,Becattini2015,PhysRevResearch.1.033058,PhysRevC.103.024903}, the transport approach~\cite{PhysRevC.96.054908,PhysRevC.97.064902,PhysRevC.98.024905,PhysRevC.99.014905,Deng:2021miw} and the non-equilibrium chiral kinetic approach~\cite{PhysRevC.96.024906,PhysRevC.99.011903,PhysRevLett.125.062301}, have successfully described the measured $\Lambda$ global spin polarization, most of them have failed to explain the measured $\Lambda$ local spin polarizations. A plausible explanation of the latter has been provided in the chiral kinetic approach through the induced quadrupole axial charge distribution in the transverse plane of a heavy ion collision~\cite{PhysRevC.99.011903,PhysRevLett.125.062301}. Also, it has recently been pointed out that adding the thermal shear contribution to that from the thermal vorticity in the fluid dynamic approach can potentially describe the measured azimuthal angle dependence of $\Lambda$ local spin polarizations~\cite{Liu:2021uhn,Becattini:2021suc,Yi:2021ryh,Liu:2021nyg}.  Indeed, the correct azimuthal angle dependence is obtained in the ``strange memory" scenario of Ref.~\cite{PhysRevLett.127.142301}, which assumes that the $\Lambda$ spin polarization is identical to its strange quark spin polarization like in the quark coalescence model for $\Lambda$ production~\cite{PhysRevC.96.024906}, after including contributions from both thermal vorticity and thermal shear.   A similar result can also be achieved in the ``isothermal local equilibrium" scenario of Ref.~\cite{Becattini:2021iol}, in which the hadronization is assumed to take place at a constant temperature to eliminate the contribution from its space-time gradients in the calculation of the thermal vorticity and shear~\cite{Becattini:2021iol}.

In almost all these theoretical studies, the $\Lambda$ spin polarization in a heavy ion collision is calculated at the end of the partonic phase and compared to experimental measurements. Since the strength of the vorticity field decreases as the hadronic matter expands~\cite{PhysRevC.94.044910}, it is of interest to study how the $\Lambda$ polarization changes during the hadronic evolution of heavy ion collisions. In this work, we study the time evolution of the global and local $\Lambda$ spin polarizations during the expansion of the hadronic matter produced in non-central Au+Au collisions at $\sqrt{s_{NN}}=19.6$ GeV by using the MUSIC+UrQMD hybrid model. Because of the lack of a dynamical description of the spin degree of freedom in the hadronic transport model, we study the dependence of $\Lambda$ spin polarizations on the temperature by assuming that they are in thermal equilibrium with the vorticity field in the expanding hadronic matter.


The paper is organized as follows: In Sec. II, we discuss the relations between the spin polarization of a spin-1/2 fermion and the thermal vorticity and shear in a hadronic matter. We then present the numerical results from our study on the global and local spin polarization of $\Lambda$ hyperons at different freeze-out temperatures for the spin degree of freedom.  Finally, a brief conclusion  and discussion is given in Sec. IV.

\section{Spin polarization of a fermion in thermal equilibrium}

The spin polarization of a fermion in a thermal medium of temperature $T$ depends on  the four-temperature vector $\beta=u/T$, where $u$ is the four flow velocity of the local medium. To  the leading order in the gradient of $\beta$, the spin polarization vector $S$ of  a fermion of four momentum $p$ at four space-time coordinate $x$ can be written as ~\cite{Liu:2021uhn,Becattini:2021suc,Yi:2021ryh,Liu:2021nyg}
\begin{eqnarray}
S^{\mu}(x,p)=&&-\frac{1}{8m}(1-n_F)\epsilon^{\mu\nu\rho\sigma}p_v\varpi_{\rho\sigma}(x)\nonumber
\\&&-\frac{1}{4m}(1-n_F)\epsilon^{\mu\nu\rho\sigma}p_v\frac{n_{\rho}p^{\lambda} \xi_{\lambda\sigma}(x)}{n\cdot p}.
\label{spin}
\end{eqnarray}
In the above,
\begin{eqnarray}\label{thermal}
\varpi_{\rho\sigma}=\frac{1}{2}(\partial_{\sigma}\beta_{\rho}-\partial_{\rho}\beta_{\sigma}),
\quad\xi_{\rho\sigma}=\frac{1}{2}(\partial_{\sigma}\beta_{\rho}+\partial_{\rho}\beta_{\sigma})
\end{eqnarray}
are the thermal vorticity and thermal shear of the medium, respectively, $n_F$ is the  Fermi-Dirac distribution function, and $n$ is a unit four vector that specifies the frame of reference. Although different forms of $S^{\mu}$ are used in Refs.~\cite{Liu:2021uhn,Becattini:2021suc,Yi:2021ryh,Liu:2021nyg}, they all become  the same if $n$ is taken to be the flow field $u$ in the fluid, which we adopt in this work. Our calculation using Eq.~(\ref{thermal}) is similar to that in Ref.~\cite{PhysRevLett.127.142301} but different from that based on the assumption of isothermal local equilibrium adopted in Ref.~\cite{Becattini:2021iol}, where  both the spatial and temporal gradients of the temperature are neglected, i.e., only contributions from the kinetic vorticity and shear are included.

By decomposing the two terms in Eq.(\ref{spin}) into the following two components,
\begin{eqnarray}
&&\boldsymbol{\varpi}_T=\frac{1}{2}\left[\boldsymbol{\nabla}\left(\frac{u^0}{T}\right)+\partial_t\left(\frac{\boldsymbol{u}}{T}\right)\right]-u^0\boldsymbol{f}+\boldsymbol{u}f^0,\nonumber
\\&&\boldsymbol{\varpi}_S=\frac{1}{2}\boldsymbol{\nabla}\times\left(\frac{\boldsymbol{u}}{T}\right)+\boldsymbol{u}\times\boldsymbol{f},
\label{omega}
\end{eqnarray}
with 
\begin{eqnarray}
&&u^0=\gamma,\quad \boldsymbol{u}=\gamma \boldsymbol{v},\nonumber
\\&&f^0=\frac{p^{\lambda}\xi_{0\lambda}}{u\cdot p},\quad \boldsymbol{f}=-(f_1,f_2,f_3),\quad f_i=\frac{p^{\lambda}\xi_{i\lambda}}{u\cdot p},
\end{eqnarray}
Eq. (\ref{spin}) can be rewritten as 
\begin{eqnarray}
&&S^0(x,p)=\frac{1}{4m}\boldsymbol{p}\cdot\boldsymbol{\varpi}_S,\nonumber
\\&&\boldsymbol{S}(x,p)=\frac{1}{4m}(E_p\boldsymbol{\varpi}_S+\boldsymbol{p}\times\boldsymbol{\varpi}_T),
\label{spinvector}
\end{eqnarray}
where $E_p=\sqrt{m^2+{\bf p}^2}$ and ${\bf p}$ are, respectively, the time and space components of the four momentum $p$.

Boosting the spin vector $S$ in the fluid frame to the particle's rest frame and using the fact that the spin  of a fermion equal to $1/2$, one finds the spin polarization ${\bf P}$ measured in experiments to be
\begin{eqnarray}
&& \boldsymbol{P}=2 \boldsymbol{S}^*=2\left[\boldsymbol{S}-\frac{ \boldsymbol{p}\times \boldsymbol{S}}{E_p(E_p+m)} \boldsymbol{p}\right].
\label{polarization}
\end{eqnarray}

\section{Numerical results}

To study the evolution of $\Lambda$ spin polarization in the hadronic phase of a heavy ion collision, we first use the MUSIC hydrodynamic model~\cite{PhysRevLett.106.042301} with the initial conditions taken from A Multi-Phase Transport (AMPT) model~\cite{PhysRevC.72.064901}  to simulate the QGP phase of the collision.  After converting the fluid elements on the freeze-out hypersurface into hadrons by the Cooper-Frye formula, we adopt the UrQMD model~\cite{Bass:1998ca,Bleicher:1999xi} to simulate the evolution of these hadrons due to their scatterings. This hybrid approach has  successfully described various soft hadronic observables, such as the charged particle yields as well as their transverse momentum spectra and flow anisotropies in relativistic heavy ion collisions at both RHIC and LHC energies~\cite{Shen:2020jwv, Zhao:2020irc,Schenke:2020mbo}. 

To determine the local temperature and flow field as well as their gradients in time and space in the hadronic matter, we use the coarse-grained method with the time step $\Delta t=0.5$ fm$/c$, and the cell size $\Delta x=\Delta y=0.5$ fm and $\Delta \eta=0.2$, where $\eta=\frac{1}{2}\ln\frac{t+z}{t-z}$ is the space-time rapidity. In this method, the temperature $T$ of a local cell is calculated from the  energy density in the cell at its  rest frame by using the  equation of state from the lattice QCD calculations~\cite{Borsanyi:2013bia}. For the flow field $\boldsymbol{v}$  in each cell, it is calculated from the average velocity of the $N$ hadrons in the cell according to
\begin{eqnarray}
\boldsymbol{v}(t,x,y,z)=\frac{\sum_i \frac{\boldsymbol{p_i}}{E_i}}{N},
\end{eqnarray}
where ${\bf p}_i$ and $E_i$ are the momentum and energy of the $i$-th hadron in the  cell.

\begin{figure}[h]
\centering
\includegraphics[width=1\linewidth]{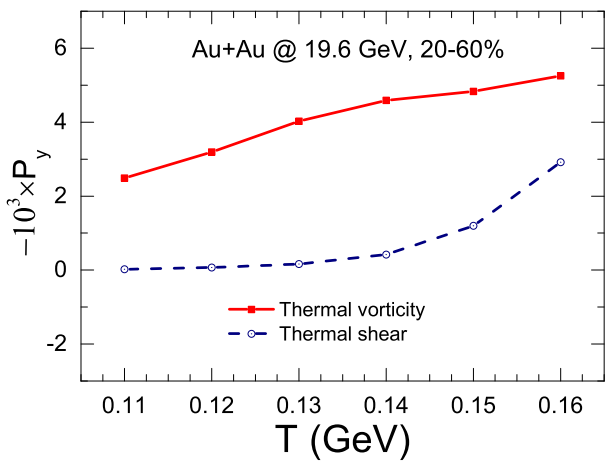}
\caption{(Color online) Spin freeze-out temperature dependence of $\Lambda$ hyperon spin polarization along the total angular momentum direction due to the thermal vorticity and the thermal shear in an expanding hadronic matter.}
\label{fig:T-Py}
\end{figure}

To illustrate how $\Lambda$ spin polarizations are affected by the temperature at which the spin degrees of freedom freeze out in relativistic heavy ion collisions, we  consider in the present study Au+Au collisions at $\sqrt{s_{NN}}=19.6$ GeV and  20-60\% centrality. As in the experimental measurements at RHIC by the STAR Collaboration~\cite{PhysRevLett.123.132301}, we only include $\Lambda$ hyperons of $|\eta|<1$ and $p_T>$0.15 GeV$/c$ in our analysis.  In Fig.~\ref{fig:T-Py}, we first show by the solid red line the global spin polarization of $\Lambda$ ($\overline{\Lambda}$) hyperons calculated from the thermal vorticity along the direction of the total angular momentum of the hadronic matter as a function of the local temperature in the hadronic matter.  It is seen that if  the spin degree of freedom  freezes out at $T_s=160$ MeV, which is usually assumed in the literature~\cite{PhysRevC.96.054908,PhysRevC.96.024906,PhysRevC.103.024903} as mentioned in the introduction, the $\Lambda$ global polarization is 5.3$\times10^{-3}$ and is close to the value in  Ref.~\cite{PhysRevC.103.024903} based on the same AMPT+MUSIC hybrid approach without the UrQMD afterburner. With decreasing temperature as the hadronic matter expands, the $\Lambda$ global spin polarization decreases  to 2.5$\times10^{-3}$ at  the kinetic freeze-out temperature of $T_s=110 $ MeV, when hadron momentum spectra and anisotropies stop changing. The above result thus shows that the $\Lambda$ global spin polarization has a strong temperature dependence in the expanding hadronic matter if it continues to be in thermal equilibrium .
 
Shown by the blue dashed line in Fig.~\ref{fig:T-Py} is the $\Lambda$ global spin polarization generated by the thermal shear in the hadronic matter, i.e., the second term in the RHS of Eq. (\ref{spin}), as a function of  local temperature. It is seen that at $T_s=160$ MeV, the $\Lambda$ global polarization due to the thermal shear is  about 2.9$\times10^{-3}$, which is comparable to that due to the thermal vorticity. We note that the total spin polarization of $\Lambda$ hyperons due to the thermal shear should have been zero without the cut of $p_T\ge 0.15$ GeV/$c$ and $|\eta|<$1 in the $\Lambda$ momentum. With decreasing spin freeze-out temperature, the $\Lambda$ global spin polarization due to the thermal shear decreases faster compared to that due to the thermal vorticity, and it becomes negligible at $T_s=140$ MeV.

\begin{figure}[h]
\centering
\includegraphics[width=1\linewidth]{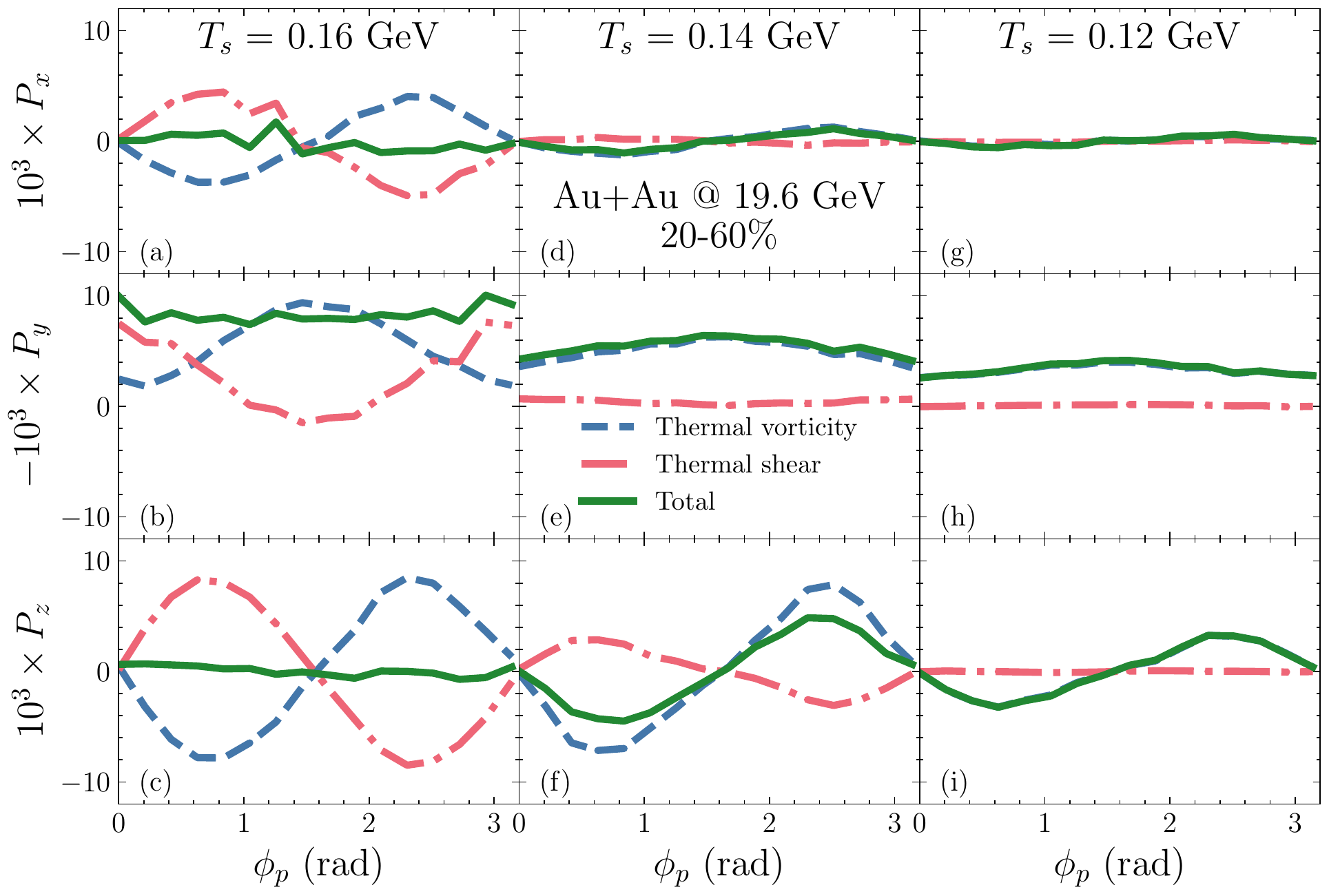}
\caption{(Color online)  The azimuthal angle dependence of the local spin polarizations $P_x, -P_y$ and $P_z$ of  $\Lambda$ hyperons generated by thermal vorticity and thermal shear at the spin freeze-out temperatures $T_s=$160, 140, and 120 MeV. }
\label{fig:Localspin}
\end{figure}

As to the local spin polarization of $\Lambda$ hyperons as a function of its azimuthal angle in the transverse plane of a heavy ion collision, it has recently been extensively studied~\cite{PhysRevLett.120.012302,PhysRevC.99.011903,PhysRevLett.125.062301,PhysRevLett.123.132301}. It is found in Refs.~\cite{PhysRevLett.127.142301,Becattini:2021iol} that both thermal vorticity and thermal shear are important in determining the local spin polarization. In the left panels (a), (b) and (c) of Fig.~\ref{fig:Localspin}, we show, respectively, the spin polarization $P_x, -P_y$ and $P_z$ of $\Lambda$ hyperons generated by the thermal vorticity (red dashed lines) and the thermal shear (blue dashed-dotted lines) as functions of the azimuthal angle $\phi_p$ of the $\Lambda$ transverse momentum if the spin degree of freedom freezes out at $T_s=160$ MeV. It is seen that $-P_y$ (left-middle panel) and $P_z$ (left-lower panel) have the form of $-\cos(2\phi_p)$ and $-\sin(2\phi_p)$, respectively, due to the thermal vorticity and $\cos(2\phi_p)$ and $\sin(2\phi_p)$, respectively, due to the thermal shear, which are similar to the findings in Refs.~\cite{PhysRevLett.127.142301,Becattini:2021iol,Yi:2021ryh}. However, such an oscillatory azimuthal angle dependence essentially disappears in the total $\Lambda$ spin polarization after adding the two contributions, which disagrees with the experimental measurements~\cite{PhysRevLett.123.132301}, unless one adopts either the ``strange memory" scenario as in Ref.~\cite{PhysRevLett.127.142301} or the ``isothermal local equilibrium" scenario as in Ref.~\cite{Becattini:2021iol}.  In the present work, we focus on the effect of hadronic evolution on the $\Lambda$ spin polarization and postpone the study of the above two scenarios for future study.

In the left-upper panel (a) of Fig.~\ref{fig:Localspin}, we also show for the first time the azimuthal angle dependence of $P_x$ generated by the thermal vorticity and thermal shear. It is seen that $P_x$ has the form of $-\sin(2\phi_p)$ due to the thermal vorticity and the form of $\sin(2\phi_p)$ due to the thermal shear. The total $P_x$ after adding the two oscillatory contributions is, however, negligible.

 \begin{figure}[h]
\centering
\includegraphics[width=1\linewidth]{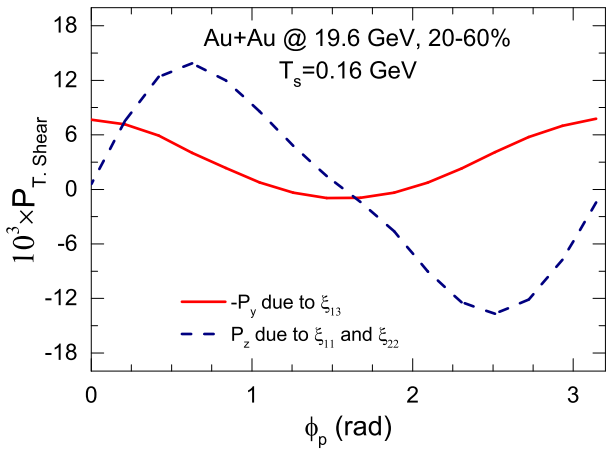}
\caption{(Color online)  Azimuthal angle dependence of $\Lambda$ spin polarization  generated by components of the thermal shear at the spin freeze-out temperature $T_s=160$ MeV.}
\label{fig:component}
\end{figure}

In Fig.~\ref{fig:component}, we further show the $\Lambda$ spin polarization generated by individual components of the thermal shear. We find that the main contribution to $-P_y$ is from the $\xi_{13}$ component and that to $P_z$ is from the $\xi_{11}$ and $\xi_{22}$ components. According to Eqs.(\ref{omega}), (\ref{spinvector}) and (\ref{polarization}), $\xi_{13}$ contributes to $-P_y$ through a term of the form $p_x^2\xi_{13}\propto (1+\cos(2\phi_p))\xi_{13}$. For $P_z$, the contribution from $\xi_{11}$ and $\xi_{22}$ is through a term of the form $p_xp_y(\xi_{11}-\xi_{22})\propto \sin(2\phi_p)(\xi_{11}-\xi_{22})$, where $(\xi_{11}-\xi_{22})$ is positive because of the positive elliptic flow of the hadronic matter in non-central Au+Au collisions at $\sqrt{s_{NN}}=19.6$ GeV.

 \begin{figure}[h]
\centering
\label{fig:T-0.14}
\end{figure}

We have also studied the azimuthal angle dependence of the spin polarization of $\Lambda$ hyperons generated by the thermal vorticity and thermal shear at a lower temperature of $T_s=140$ MeV during the later stage of the hadronic evolution by assuming that the spin degree of freedom remains in thermal equilibrium. As shown by the red dashed lines in the middle panels (d), (e) and (f) of Fig.~\ref{fig:Localspin}, $P_x, -P_y$ and $P_z$  generated by the thermal vorticity at this temperature have the same azimuthal angle dependence as those at $T_s=160$ MeV. Although $P_x$ and  $-P_y$ become smaller at the lower temperature, $P_z$ does not change much with temperature. For the contribution from the thermal shear at $T_s=140$ MeV,  $P_x$ and $-P_y$ become negligible, while $P_z$ decreases by a factor of 3 as shown by the blue dashed-dotted lines. By adding contributions from both  thermal vorticity and thermal shear, $P_x$ and $P_z$ have the form of $-\sin(2\phi_p)$ and $P_y$ has the form of $-\cos(2\phi_p)$. These results suggest that the $\Lambda$ local spin polarization depends strongly on its evolution during the hadronic phase of heavy ion collisions.

 \begin{figure}[h]
\centering
\label{fig:T-0.12}
\end{figure}

Finally, we consider an even lower temperature of $T_s=120$ MeV, corresponding approximately to the kinetic freeze-out of the expanding hadronic matter, and the results are shown in the right panel{s (g), (h) and (k) of Fig.~\ref{fig:Localspin}. It is seen from the red dashed lines that the $\Lambda$ local spin polarizations $P_x, -P_y$ and $P_z$ generated by the thermal vorticity become even smaller, although they have the same $\phi_p$ dependence as that at higher temperatures. However, for the $\Lambda$ spin polarization generated by the thermal shear, its values along all $x$, $y$ or $z$ directions become negligibly small. Thus, the total $P_x$, $-P_y$ and $P_z$ have the same $\phi_p$ dependence as that at $T_s=140$ MeV, albeit somewhat smaller.

\section{Conclusion and Discussion}

Using the AMPT model  initial conditions to the MUSIC hydrodynamic model for the QGP phase, which is followed by the UrQMD model for the hadronic phase  in  relativistic heavy ion collisions, we have studied in this paper the global and local spin polarizations of  $\Lambda$ hyperons as functions of the freeze-out temperature of the spin degree of freedom in the hadronic phase. We have found that both the $\Lambda$ global and local spin polarizations due to the thermal vorticity decrease by a factor of 2 if the spin freeze-out temperature drops from 160 to 140 MeV, while those due to the thermal shear already becomes negligibly small at temperature equal to 140 MeV. This result suggests the importance of including in theoretical studies the dynamical evolution of the spin degrees of freedom and its freeze-out in the expanding hadronic matter. Such a study requires a hadronic transport model that takes  into account explicitly the spin degrees of freedom of hadrons, such as in Refs.~\cite{Xu:2012hh,Xia:2014qva,Xia:2014rua,PhysRevD.104.016022,PhysRevD.104.016029} on the nucleon spin transport in heavy ion reactions at lower energies, which we plan to study in the future.

\section*{ACKNOWLEDGEMENTS}
We thank Huichao Song for helpful discussions and suggestions.
This work was supported by INFN-SIM national project and linea di intervento 2 for HQCDyn at DFA-Unict (Y.S.), the National Natural Science Foundation of China under Grant No. 11905302 (Z.Z.), the U.S. Department of Energy under Contract No. DE-SC0015266 and the Welch Foundation under Grant No. A-1358 (C.M.K.), and the U.S. National Science Foundation (NSF) under grant numbers ACI-2004571 (W.Z.).

\bibliography{ref.bib}
\end{document}